\documentclass[11pt]{article}
\usepackage{blois,epsfig}
\usepackage[latin1]{inputenc}
\usepackage[OT1]{fontenc}

\def\Journal#1#2#3#4{{#1} {\bf #2}, #3 (#4)}

\def\NIM{\em Nucl. Instrum. Methods}

\def\PRL{\em Phys. Rev. Lett.}

\def\APH{\em Astropart. Phys.}
\def\AJ{\em Astrophys. J.}
\def\JPH{\em J. Phys. G}

\def\be{\begin{equation}}
\def\ee{\end{equation}}
\def\bea{\begin{eqnarray}}
\def\eea{\end{eqnarray}}

\begin{document}
\vspace*{4cm}
\title{Mass Composition Studies of the Highest Energy Cosmic Rays}

\author{Jos\'e Bellido, for the Pierre Auger Collaboration}

\address{Department of Physics, The Pennsylvania State University, USA~\footnote{Currently at: Physics Department, The University of Adelaide, SA-5005, Australia}}

\maketitle\abstracts{The determination of the mass composition of the highest energy cosmic rays is one of the greatest challenges in cosmic ray experiments. The highest energy cosmic rays are only detected indirectly because of their very low flux. Using the atmosphere as a large target, Air Fluorescence Detectors are capable of tracing the evolution of the size of the Extensive Air Shower through the atmosphere (the shower longitudinal profile). The analysis of the characteristics of the detected longitudinal profiles is currently the most reliable way for extracting some information about the primary cosmic ray mass composition. In this proceeding, I will describe in some detail the Pierre Auger elongation rate studies, and I will show the potential for mass composition studies using the surface and the fluorescence detectors information as part of a single analysis. The interpretation of the current data with regard to mass composition, relies heavily on high energy hadron interaction models. Using standard hadron interaction models, the data suggest that the composition becomes lighter up to about 2 $\times$ 10$^{18}$ $eV$ and above that it becomes heavier again. This apparent change in the mass composition at 2 $\times$ 10$^{18}$  $eV$ seems to be correlated with a spectrum index change in the observed energy spectrum.}

\section{Introduction}\label{subsec:intro}

 Energetic cosmic rays interacting with the atmosphere initiate a cascade of secondary particles known as an Extensive Air Shower (EAS) that travels near the speed of light. The number of particles existing in the EAS changes as it evolves throught the atmosphere. The EAS can be considered to be composed of hadrons, muons and an electromagnetic component. Each of these components evolve differently through the atmosphere.

The number of particles in the EAS (or shower size) as a function of the atmosphere slant depth is called the shower longitudinal profile. Most of the EAS energy is dissipated through the electromagnetic component (where $N_{e}$ is the number of $e^{\pm}$). Therefore, the shower size ($N_{e}$) increases until the average energy of the $e^{\pm}$  in the EAS is about the critical energy ($E_{c} = 81$ $MeV$). The critical energy is when the rate of energy loss due to collisions and ionization begins to exceed that due to radiation. This happen when the energy of  $e^{\pm}$ is smaller than the energy lost by ionization (2.2 $MeV/g/cm^{2}$) after traveling one radiation length (37 $g/cm^{2}$ in air). A radiation length is the grammage path length required so that the energy of  $e^{\pm}$  is attenuated by the factor $1/e$ due to bremsstrahlung radiation. Photons attenuate to $1/e$ due to pair production within a similar grammage path length\cite{sommers}.

 The slant depth at which the longitudinal profile reaches its maximum is called $X_{max}$ and this is the main shower parameter used to extract information about the mass composition of the primary cosmic ray.

\begin{figure}[ht]
\begin{center}
\center {\epsfig{file=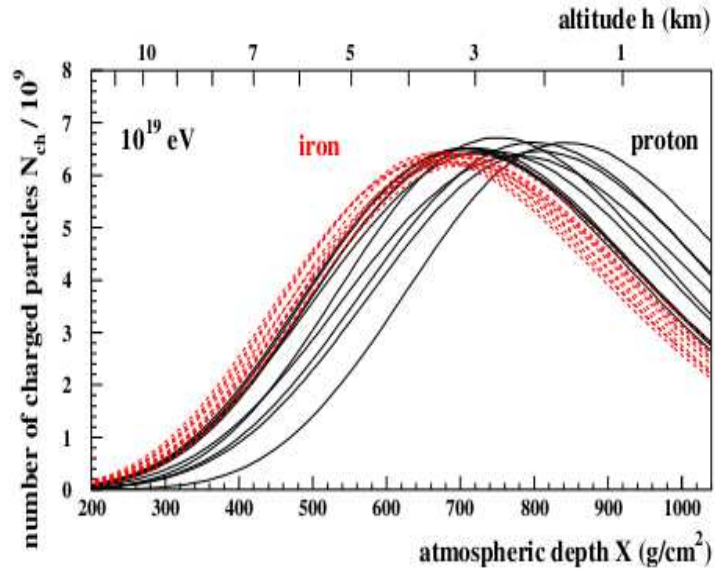,width=10cm,height=8cm}}
\caption {Individual longitudinal shower profiles for $10^{19}$ $eV$ proton (solid black lines) and iron (dashed red lines) vertical showers according to MC simulations$^2$.\label{Simulation_Fe_P} }

\end{center}
\end{figure}

 The EAS initiated by a higher energy cosmic ray has more energy to dissipate, so it takes longer for  the particles to reach the critical energy. Therefore, for a given composition, the average $X_{max}$ should increase as a function of the cosmic ray energy. On the other hand, heavier nuclei can be considered as the superposition of many nucleons. For an iron nucleus with energy E, each nucleon would have energy E/56,  and the superposition of 56 lower energy subshowers will reach $X_{max}$ earlier (higher in the atmosphere). In the superposition model, because each nucleon has a smaller energy, secondary pions reach sooner (than a single nucleon with energy E) to lower energies where they can decay to muons instead of transfering energy to the electromagnetic cascade\cite{sommers}. Then heavier nuclei will produce more muons than lighter nuclei and they will also reach $X_{max}$ higher in the atmosphere (smaller $X_{max}$).  Figure \ref{Simulation_Fe_P} shows the expected shower profiles for proton and iron showers according to MC simulations.

\begin{figure}[ht]
\begin{center}
\center {\epsfig{file=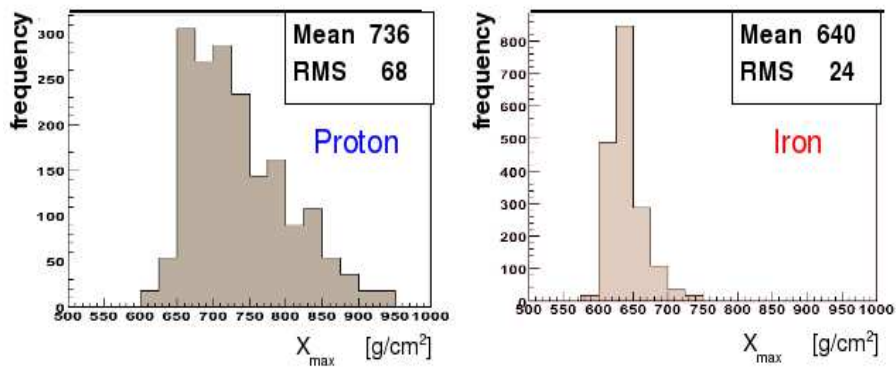,width=12cm,height=5cm}}
\caption {Expected distributions for  $X_{max}$  for $10^{18}$ $eV$ proton and iron showers according to MC simulations using the QGSJET model for high energy hadron interactions.  \label{XmaxFluctuations}}
\end{center}
\end{figure}

 Proton and iron showers not only have their average $X_{max}$ values at somewhat different slant depths, but also their $X_{max}$ fluctuations are different. Figure \ref{XmaxFluctuations} shows the distribution of $X_{max}$ for fixed-energy proton and iron showers according to MC simulations. Iron showers have a norrower distribution.

  Experimentally, there are some issues when measuring the muon abundance. The muon abundance is only measured at ground level (where the ground particle detectors are deployed). The relative abundance of muons at the ground is therefore also dependent on the particular zenith angle of the shower. Another issue is that, depending on the relative location of each ground particle detector, they will sample the muon content at different ages of the shower development (except for vertical showers). Furthermore, efficiently separating  the muon content from the electromagnetic signal requires more complex detector systems (i.e. detectors buried underground). However, fluorescence detectors have the capability to track the shower development through the atmosphere and $X_{max}$ is measured with good accuracy. In this proceeding I will focus on mass composition studies based on $X_{max}$ observations with the fluorescence detector (FD) of the Pierre Auger Observatory.

\section{Shower Reconstruction and Resolution on $X_{max}$}

\begin{figure}[ht]
\begin{center}
\center {\epsfig{file=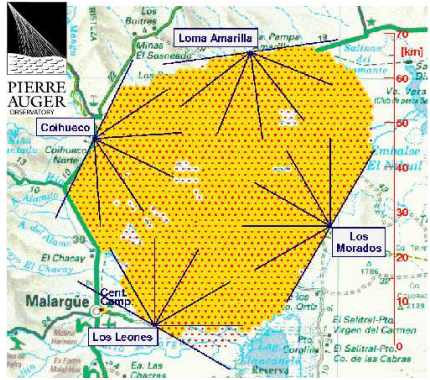,height=5cm}}
\caption {Layout of the Pierre Auger Observatory.   \label{pao}}
\end{center}
\end{figure}

The Pierre Auger Observatory is a hybrid cosmic ray detector \cite{pao}. It uses ground particle detectors to sample the shower cascade particles at the ground and also fluorescence detectors to track the shower development through the atmosphere. Figure \ref{pao} shows the layout of the the ground detectors (or stations) and the layout of the fluorescence detectors . There are 1600 stations spaced by 1.5 $km$ and covering a total area of about 3000 $km^{2}$. There are also four fluorescence detectors (Los Leones, Los Morados, Loma Amarilla and Coihueco) located at the edges of the array and overlooking the array as indicated in figure \ref{pao}.     

Each fluorescence detector is a building that contains six telescopes. Each telescope uses a spherical mirror to focus the light onto a camera. The camera is formed by 440 hexagonal PMTs (or pixels) and the angular size of each pixel is 1.5$^\circ$. The total field of view of the camera is approximately 28$^\circ$ elevation by 30$^\circ$ azimuth \cite{pao}.   

\begin{figure}[ht]
\begin{center}
\center {\epsfig{file=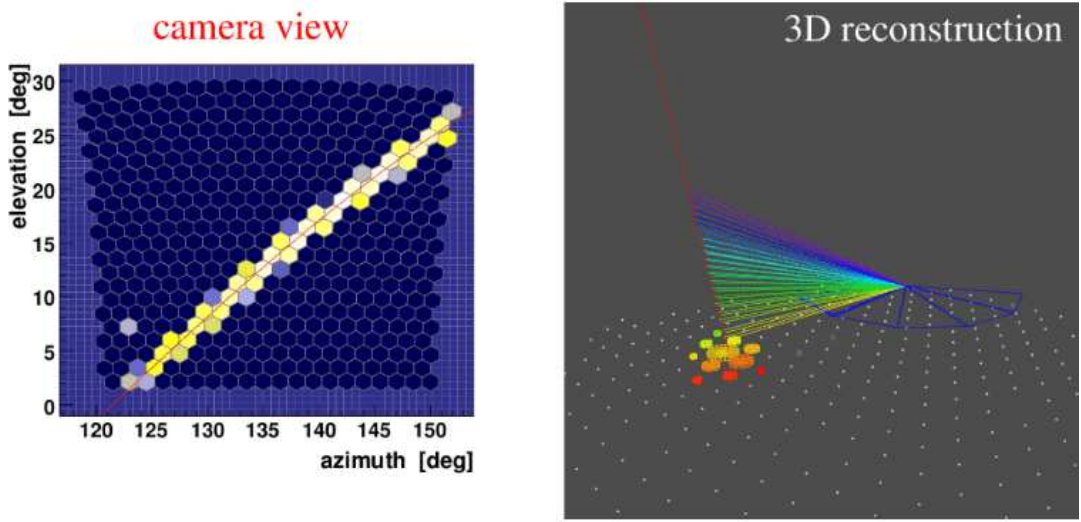,height=7cm}}
\caption {On the left: Example of a Cosmic Ray seen by the Fluorescence Detector. On the right: The same event seen in 3D. The dots denote the location of the surface detectors and the larger circles are for the stations that were triggered by this event.   \label{camera}}
\end{center}
\end{figure}

 Figure \ref{camera} shows an example of a EAS seen by the fluorescence detector and also detected  by the ground array (hybrid event). The figure on the left (fig. \ref{camera}) shows the shower track crossing through the camera field of view. The figure on the right shows the shower axis landing on the surface array, the larger circles show the stations triggered by the shower. 

 The uncertainty in the reconstructed shower geometry is much smaller for hybrid events. The time and location  information from the station closest to the shower core is a powerful constraint on the reconstructed shower geometry. On average, the uncertainty in the hybrid reconstructed arrived direction is about 0.6$^\circ$ \cite{bonifazi}. For mass composition studies we use only hybrid events (events that triggered only the FD have a poorer geometry resolution, and therefore, poorer $X_{max}$ resolution).

\begin{figure}[ht]
\begin{center}
\center {\epsfig{file=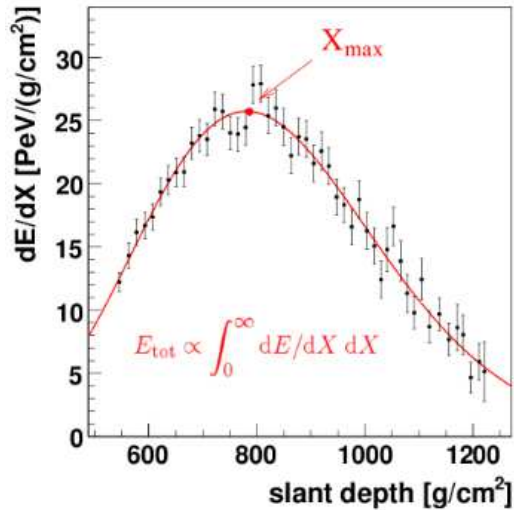,height=7cm}}
\caption {Reconstruction of the shower profile from the event seen in figure \ref{camera}.   \label{profile}}
\end{center}
\end{figure}

 Once the shower geometry is determined, the shower longitudinal profile is estimated. The longitudinal profile can be estimated as $N_{e}$ (shower size) or as $dE/dX$ (deposited energy) as a function of slant depth. In the Pierre Auger Observatory we estimate    $dE/dX$ as a function of slant depth \cite{MU-profile}. Figure \ref{profile} shows the reconstructed longitudinal profile for the same event shown in figure \ref{camera}. The integral of  $dE/dX$ provides the total energy  that is deposited in the atmosphere ($E_{tot}$), and the total energy contained by the primary cosmic ray is obtained after a correction to account for the missing energy (energy that is not dissipated in the atmosphere). The missing energy correction is both model dependant and primary mass composition dependant. The range of this correction is about 10\% \cite{barbosa,pierog}.

\begin{figure}[ht]
\begin{center}
\center {\epsfig{file=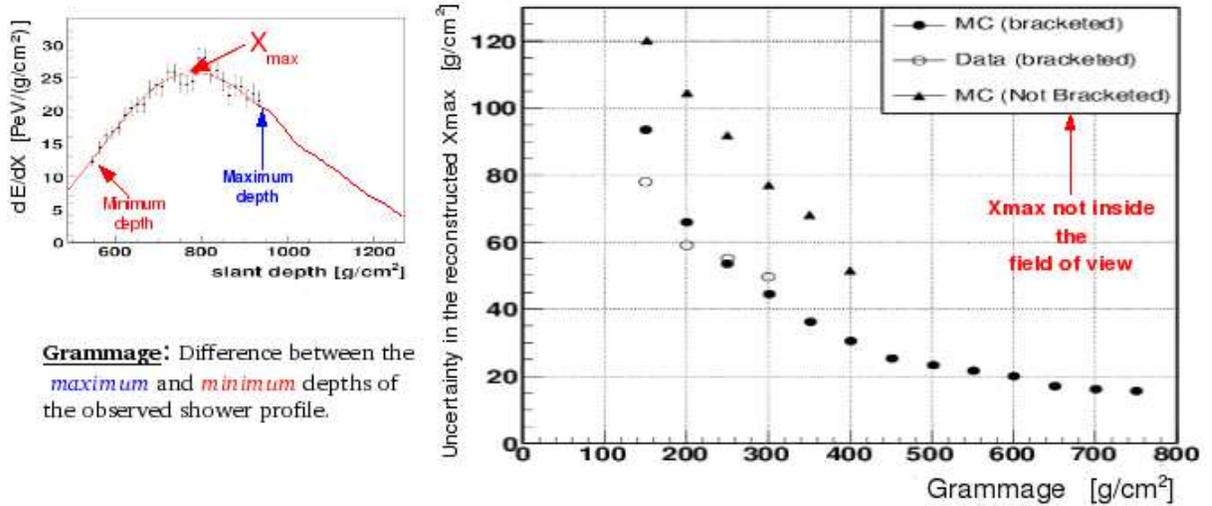,height=7cm}}
\caption {Estimated resolution on the reconstructed $X_{max}$ as a function of the observed grammage length (plot on the right).  The plot on the left shows the start (minimum depth) and the end (maximum depth) of the fraction of the shower profile seen by the FD (region with data points).  \label{resolution}}
\end{center}
\end{figure}

 The function used to fit the observed longitudinal profile (red line in figure \ref{profile}) is the Gaisser-Hillas function \cite{gaisser-hillas}, shown in equation \ref{GH}. It has four free parameters: $X_{max}$, $(dE/dX)_{max}$ (or $N_{max}$ when the longitudinal profile is in terms of $N_{e}$), $\lambda$ and $X_{0}$. $X_{max}$ is the slant depth where the shower reached its maximum size, and $(dE/dX)_{max}$ is the energy deposited at $X_{max}$. The other two parameters  $\lambda$ and  $X_{0}$, are shape parameters. $\lambda$ relates to the width of the profile and $X_{0}$ to the start of the profile. Even though $X_{0}$ relates to the start of the profile, it is not to be interpreted as the depth of the first interaction, since it often takes negative values when fitting simulated showers and also real showers. 
 
\begin{equation}
dE/dX = (dE/dX)_{max} \left( \frac{X-X_{0}}{X_{max}-X_{0}}\right)^{\frac{X_{max}-X_{0}}{\lambda}} e^{\frac{X_{max}-X_{0}}{\lambda}}
\label{GH}
\end{equation}
 The uncertainty in the reconstructed  $X_{max}$ in a given event depends on how large a fraction of the profile is seen by the FD, and whether $X_{max}$ is inside this fraction of the profile. We use MC events to estimate the uncertainty in the reconstructed $X_{max}$ as a function of the observed grammage length. Figure \ref{resolution} (plot on the right) shows  the uncertainties for showers with  $X_{max}$ bracketed\footnote{$X_{max}$ bracketed means that the reconstructed $X_{max}$ is within the fraction of the profile seen by the FD} (solid circles), and for not bracketed showers (solid triangles). For a cross check of the $X_{max}$ uncertainty estimates we selected real showers where the entire profile is seen by the FD (like the one shown in figure \ref{profile}). The uncertainty in the reconstructed $X_{max}$ is small ($<$ 20 $g/cm^{2}$) for such showers. This reconstructed  $X_{max}$  is considered as the true value, a new  $X_{max}$ reconstruction is performed (basically a new fit to the Gaisser-Hillas function) but this time using only a smaller fraction of the observed profile, but still with $X_{max}$ within this fraction (left plot in figure \ref{resolution}). The uncertainty of the reconstructed   $X_{max}$ estimated using real showers is shown with open circles in figure \ref{resolution}. It is consistent with the    $X_{max}$ uncertainty estimated using MC events.

 For mass composition analysis we want to use only showers where $X_{max}$ is reconstructed with an uncertainty smaller than 40 $g/cm^{2}$. This small uncertainty  is achieved by rejecting showers with grammage length smaller than 320 $g/cm^{2}$ and showers  with $X_{max}$ not being bracketed. The average uncertainty in the reconstructed  $X_{max}$ is about 20  $g/cm^{2}$ when applying these cuts. High energy showers more often land further away from the FD (compared with low energy showers) and because of the FD limited field of view (see fig. \ref{XmaxBias}) they have larger observed grammage lengths on average. Thefore, the average uncertainty in their  reconstructed  $X_{max}$ is slightly smaller (than those for lower energy showers).

\section{The Mean $X_{max}$ and the FD limited Field of View}

\begin{figure}[ht]
\begin{center}
\center {\epsfig{file=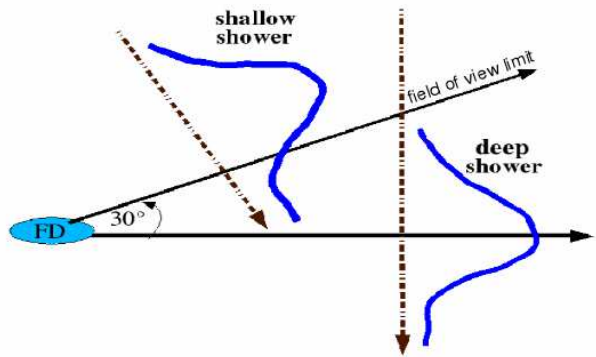,height=5cm}}
\caption {Diagram showing the possible bias in the estimated $\langle X_{max} \rangle$ due to the limitted field of view of the Auger fluorescence detector.   \label{XmaxBias}}
\end{center}
\end{figure}

 The FD has a limited field of view in elevation ranging from about 2$^\circ$ to 30$^\circ$, introducing a bias in the distribution of  $X_{max}$ from the observed showers. This bias is amplified by demanding that $X_{max}$ be within the observed profile ( $X_{max}$ bracketed). The reason for this bias in the  $X_{max}$ distribution is because many showers landing close to the FD will have their  $X_{max}$ outside (above) the field of view (see fig \ref{XmaxBias}) and the observed profile will not have  a bracketed  $X_{max}$ or the shower will simply not be detected. As a result the mean  $X_{max}$ will appear to be larger (deeper). A similar bias happen for high energy showers. High energy showers develop their   $X_{max}$ deeper in the atmosphere, then for some vertical (or near vertical) showers  $X_{max}$ will be below the ground (see fig \ref{XmaxBias}), therefore rejected from the analysis. In this case the mean  $X_{max}$ appears to be smaller (shallow).

\begin{figure}[ht]
\begin{center}
\center {\epsfig{file=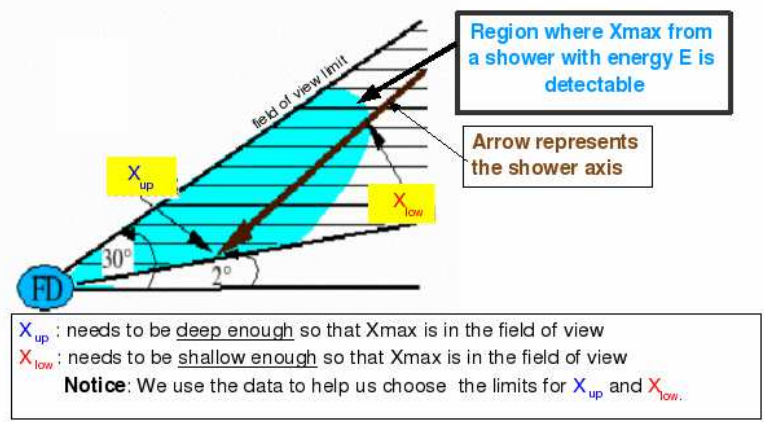,height=6cm}}
\caption {Diagram showing the definition of $X_{low}$ and $X_{up}$. $X_{low}$ and $X_{up}$ are the lower and upper limits of the slant depth along the shower axis that is inside the FD field of view.   \label{XlowXupDiagram}}
\end{center}
\end{figure}

In order to avoid the bias in the estimated mean  $X_{max}$, showers with specific geometries relative to the FD are rejected. To identify the optimum shower geometries to be used for determining the mean  $X_{max}$ ($ \langle X_{max} \rangle$) values, we introduced the parameters $X_{up}$ and $X_{low}$. These parameters are the lower and upper limits of the slant depth along the shower axis that is inside the FD field of view. These limits may be defined at where the shower axis intercepts the FD field of view limit or where the shower axis intercepts the maximum distance that a shower with energy E is still detectable (as shown in figure \ref{XlowXupDiagram}).

\begin{figure}[ht]
\begin{center}
\center {\epsfig{file=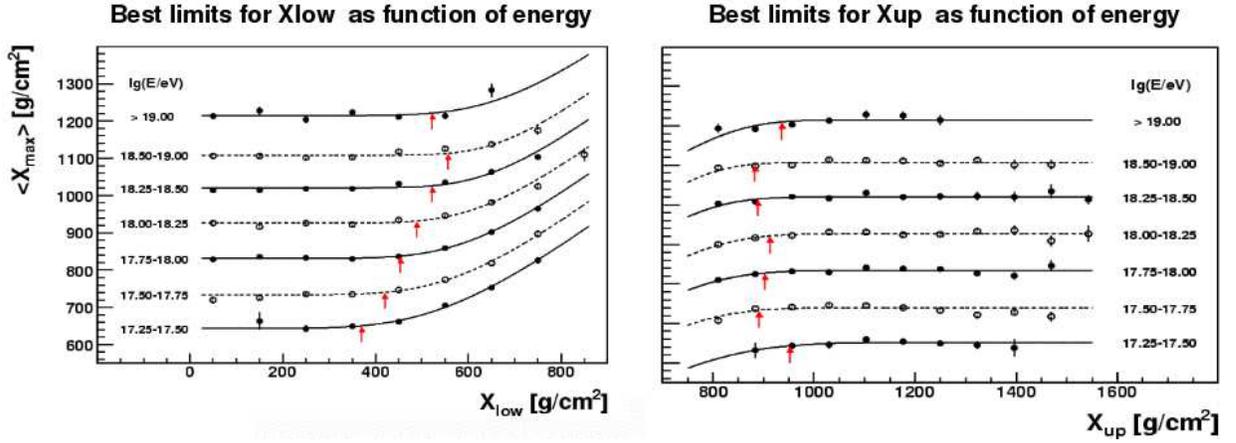,height=6cm}}
\caption {The mean $X_{max}$ values are plotted as a function of $X_{low}$ and $X_{up}$ for each energy range. The limits for the accepted  $X_{low}$ and $X_{up}$ values (at each energy range) correspond to the values where the fitted functions start to deviate from being flat. For didactic purposes, the  mean $X_{max}$ values are offset by 75 $g/cm^{2}$ consecutively.  \label{XlowXupCuts}}
\end{center}
\end{figure}

The basic idea is that  $X_{low}$ has to be shallow (small) enough, and  $X_{up}$ has to be deep (large) enough to guarantee that  $X_{max}$ will be within the field of view. Notice that the values for  $X_{low}$ and   $X_{up}$ depend only on the geometry of the shower and not on the particular depth of the shower profile (but it does also depend in the overall distribution of shower profile depths). In order to determine the optimum ranges for  $X_{low}$ and   $X_{up}$ as a function of energy  we use real data. Figure \ref{XlowXupCuts} shows the mean  $X_{max}$ as a function of  $X_{low}$ (plot on the left) and   $X_{up}$ (plot on the right) for different energies. The arrows in figure \ref{XlowXupCuts} indicates where the fitted functions  start to deviate from being flat and this defines the limit values for   $X_{low}$ and   $X_{up}$. The flat region is the unbiased one.

\section{Checking the Anti-Bias Cuts with MC Data}

 It is possible to check the performance of the cuts applied to $X_{low}$ and $X_{up}$ using MC data. We have simulated MC showers and pass these showers through the detector simulation, where the FD and the SD components are simulated in detail \cite{MC-simulation,prado}. We then reconstruct all triggered MC showers in the same way as real showers. The input compositions for the MC simulation are pure iron, pure proton, and a mixed composition of 50\% iron and 50\% proton.

\begin{figure}[ht]
\begin{center}
\center {\epsfig{file=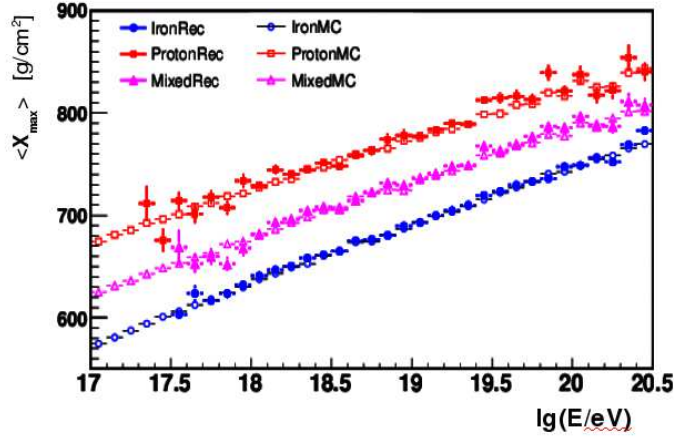,height=6cm}}
\caption {Reconstructed mean  $X_{max}$ as a function of energy for MC generated data (solid symbols). The open symbols indicate the true mean   $X_{max}$ from the MC input. The simulated primary compositions are proton (squares), iron (circles) and a proton/iron mixed composition (triangles).    \label{MC_ER}}
\end{center}
\end{figure}

 Figure \ref{MC_ER} shows that the reconstructed mean   $X_{max}$ are consistent with the  mean   $X_{max}$ from MC inputs. This indicates that the cuts applied on  $X_{low}$ and $X_{up}$ efficiently removed the bias in  $X_{max}$. The minimum energy where the mean   $X_{max}$ can still be measured (without being biased) is just above 10$^{17.5}$ $eV$. The limits for $X_{low}$ and $X_{up}$ are not the same for all compositions. For each composition we need to follow the procedure described in the previous section. This is because the limits  for $X_{low}$ and $X_{up}$ depend on the particular range of the observed (or simulated)  $X_{max}$ values.

\begin{figure}[ht]
\begin{center}
\center {\epsfig{file=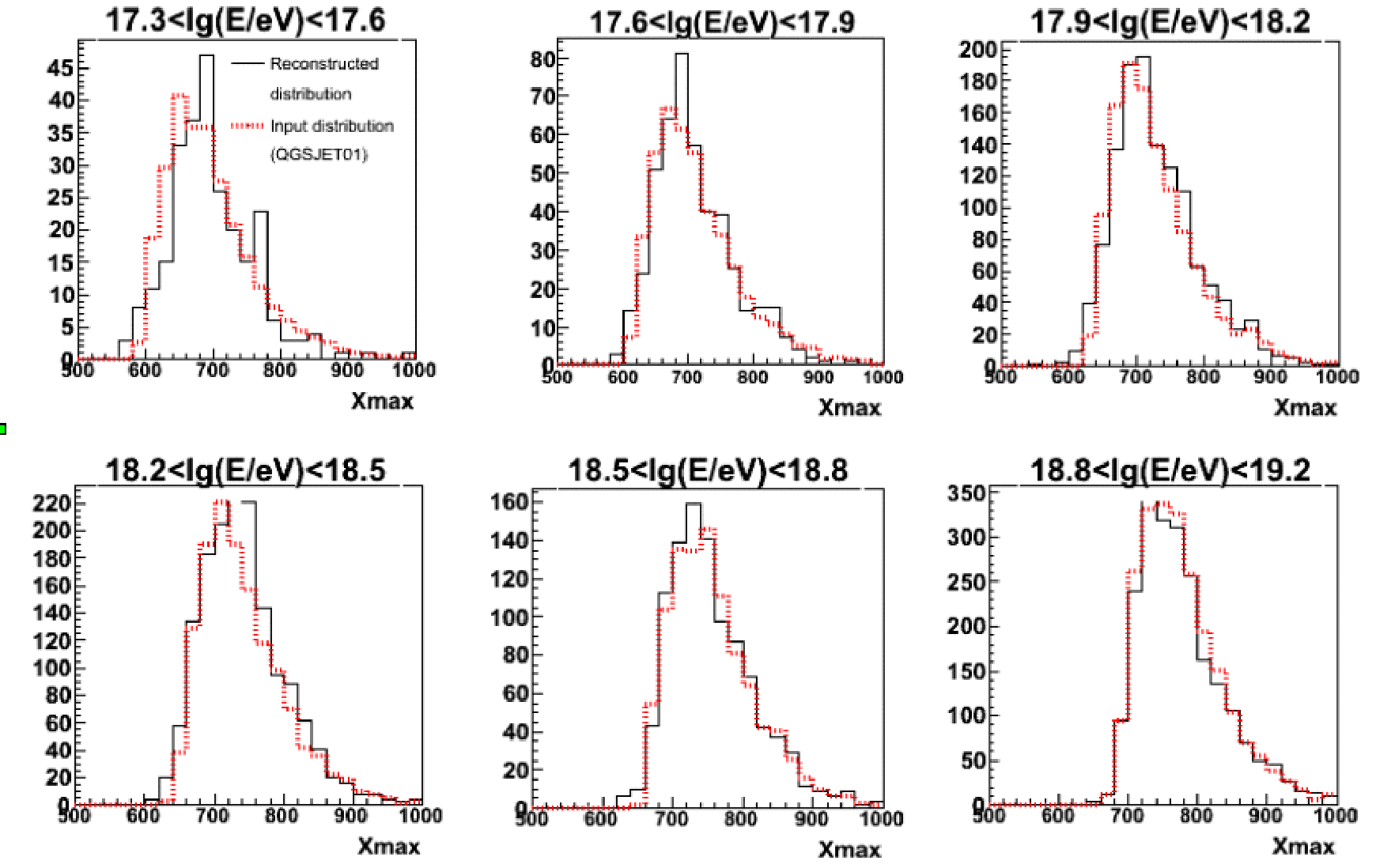,height=6cm}}
\caption {Distributions for {\bf proton} showers at different energy ranges (see caption in figure \ref{MC_Fe}).  \label{MC_P}}
\end{center}
\end{figure}

\begin{figure}[ht]
\begin{center}
\center {\epsfig{file=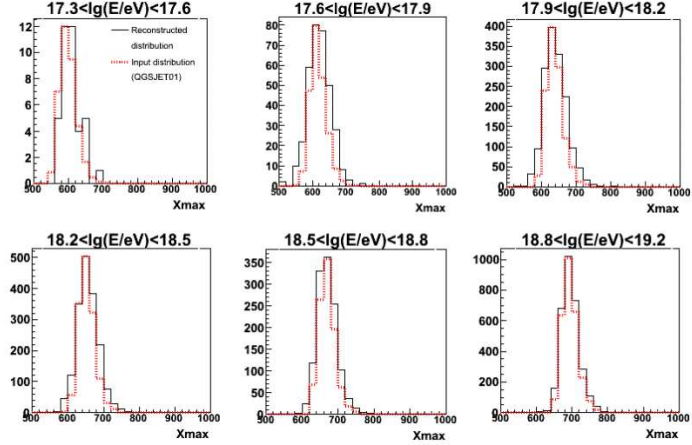,height=6cm}}
\caption {Distribution for {\bf iron} showers at different energy ranges according to MC simulations. The solid (black) lines show the reconstructed  $X_{max}$ distributions, and the dotted (red) lines show the MC input  $X_{max}$ distributions (using to the QGSJET01 model).     \label{MC_Fe}}
\end{center}
\end{figure}

We also confirm the stability of the $X_{max}$ distributions with respect the quality cuts and the cuts on $X_{low}$ and $X_{up}$.   This is of particular interest since the fluctuations in $X_{max}$ are sensitive to primary mass composition. Figures \ref{MC_P} and \ref{MC_Fe} show the distributions for  $X_{max}$ for proton and iron primaries respectively. The solid lines indicate the reconstructed distributions and the dotted lines indicate the MC input distributions, both distributions are consistent, so the quality cuts and the  cuts on  $X_{low}$ and $X_{up}$ are not biasing the $X_{max}$ distributions.

\section{Results and Discussion}

\begin{figure}[ht]
\begin{center}
%\center {\epsfig{file=ER_result.ps,height=5cm}}
\center {\epsfig{file=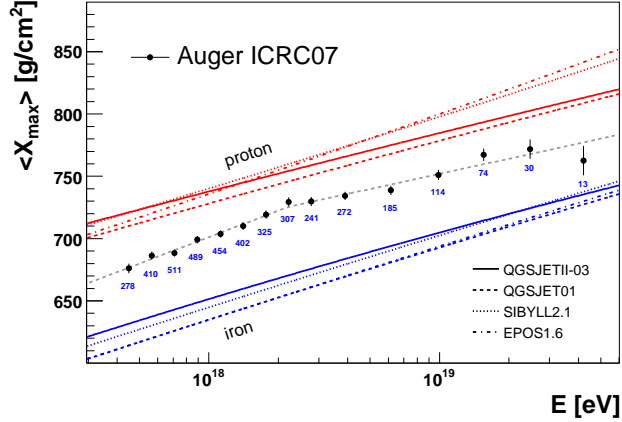,height=6cm}}
\caption { Auger results for the Mean $X_{max}$ measurements as a function of energy$^{11}$.    \label{ER_icrc}}
\end{center}
\end{figure}

 Figure \ref{ER_icrc} shows the reconstructed mean   $X_{max}$  as a function of energy as measured with the Auger  data \cite{MU-icrc}.  The blue and red lines are the expected  mean  $X_{max}$ values for iron and proton showers respectively for different hadronic interaction models.

\begin{figure}[ht]
\begin{center}
\center {\epsfig{file=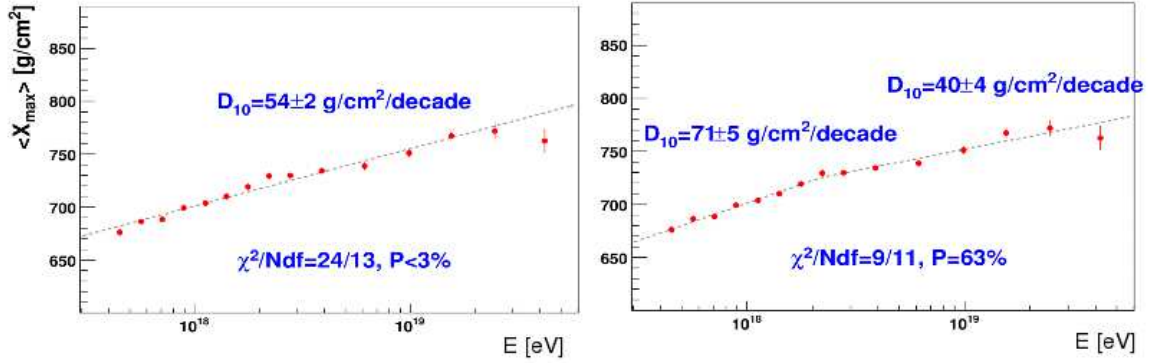,height=5cm}}
\caption { The plot on the left shows a fit to a single line. The plot on the right shows a fit to a broken line$^{11}$. \label{ER_line}}
\end{center}
\end{figure}

The rate that the mean  $X_{max}$ increases per decade in  energy is known as the \emph {elongation rate}. For a pure composition the expected elongation rate is model dependent and it is about 50 $g/cm^{2}/decade$ (as shown with blue and red lines in figure \ref{ER_icrc}).  The plot on the left in figure \ref{ER_line} shows a linear fit to the data ($\langle X_{max} \rangle =A+D_{10}\times lgE$). The slope of this fit ($D_{10}$) corresponds to the measurement of the elongation rate ($D_{10} = 54 \pm 2$ $g/cm^{2}/decade$). However, the $\chi^2/Ndf$ of the fit  (24/13) suggest that a straight line may not be the best fit for the entire energy range.  The probability that the data will follow a single line is smaller than 3\%. The plot on the right in figure  \ref{ER_line} shows that a broken line fits better the data. The probability that the data follows a broken line is 63\%. The break point according to the fit is at about 2 $EeV$. The elongation rate below and above  2 $EeV$ is 71 $ \pm 5$ $g/cm^{2}/decade$ and  40 $ \pm 4$ $g/cm^{2}/decade$ respectively \cite{MU-icrc}. The uncertainties in the elongation rate measurements come from the fit and do not take into account the systematic uncertainties on the measured mean  $X_{max}$ values. The systematic uncertainties in the estimated mean $X_{max}$ are larger at lower energies. Below 1 $EeV$ the systematic uncertainties are smaller than  15  $g/cm^{2}$ and above 1 $EeV$ they are smaller than 12 $g/cm^{2}$.

If the hadron interaction models were entirely correct, we would interpret the large elongation rate result for the energy region below 2 $EeV$ (71 $ \pm 5$ $g/cm^{2}/decade$) as a mixed composition that is becoming lighter with energy, and the small  elongation rate result for the energy region above 2 $EeV$ (40 $ \pm 4$ $g/cm^{2}/decade$) as a mixed composition that is becoming heavier with energy. However, this sudden change in the elongation rate at 2 $EeV$ may be the result of a high energy hadronic interaction property (instead of a composition change). 

\begin{figure}[ht]
\begin{center}
\center {\epsfig{file=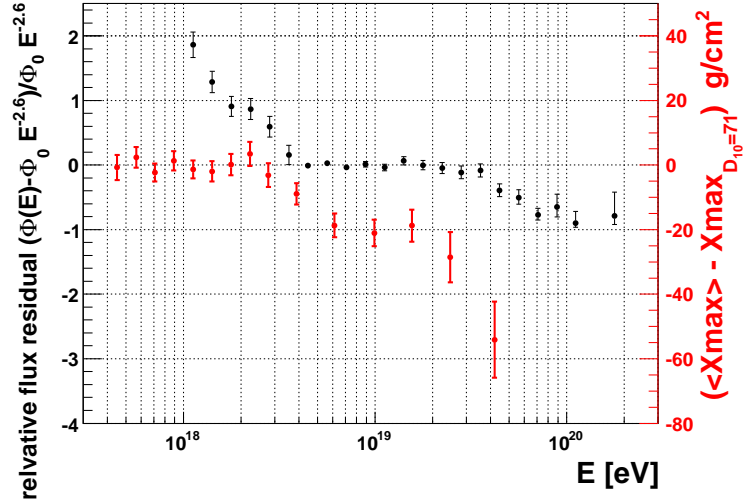,height=7cm}}
\caption {Comparing the features observed in the energy spectrum$^{12}$ against the features observed in the elongation rate measurements$^{11}$.   \label{ER_spectrum}}
\end{center}
\end{figure}

Interestingly the energy spectrum presents a change of spectral index at about 2 $EeV$. This feature in the energy spectrum is known as the ankle. Figure \ref{ER_spectrum} shows the residuals of the measured cosmic ray flux with respect to a cosmic ray flux with a spectrum of the form $ \propto E^{-2.6}$ (black symbols). A spectral index of 2.6 corresponds to the one measured  between 2$\times$10$^{18}$ $eV$ and 3$\times$10$^{19}$ $eV$ \cite{roth}, that is why the residuals curve is flat within this region. Outside these energy ranges the measured cosmic ray energy spectrum has a larger spectral index. In the same plot (fig. \ref{ER_spectrum}) we show (with red symbols) the residuals of the mean $X_{max}$ with respect to a linear function (in lgE space) that corresponds to the best fit to the elongation rate  data below 2 $EeV$ (see figure \ref{ER_line}, plot on the right). Notice that both residual plots are consistent with a feature at 2 $EeV$. Unfortunately there is not yet enough hybrid data to see whether the elongation rate plot has another feature at 3$\times$10$^{19}$ as the energy spectrum does.

 The consistency between the features observed in the  energy spectrum and in the elongation rate at 2 $EeV$ suggests that both features have the same origin. Unfortunately, this evidence is not enough to distinguish between a mass composition change or a hadron interaction physics change. However, the consistent features in the energy spectrum and in the elongation rate give us some confidence that the feature observed in the elongation rate is a real effect and not an artifact due to any detection or reconstruction bias. 

 So far the Pierre Auger collaboration has only presented mass composition studies using the mean $X_{max}$ information. New studies of the elongation rate and also studies of the  $X_{max}$ fluctuations should be released soon. As already mentioned earlier in this proceeding, the fluctuations on $X_{max}$ are also sensitive to mass composition. In this new study, the systematic uncertainties on the measured mean $X_{max}$ values may be slightly reduced and the statistics will be larger, allowing us to go up in energy to about 5$\times$10$^{19}$ $eV$ with small statistical uncertainties.  

\begin{figure}[ht]
\begin{center}
\center {\epsfig{file=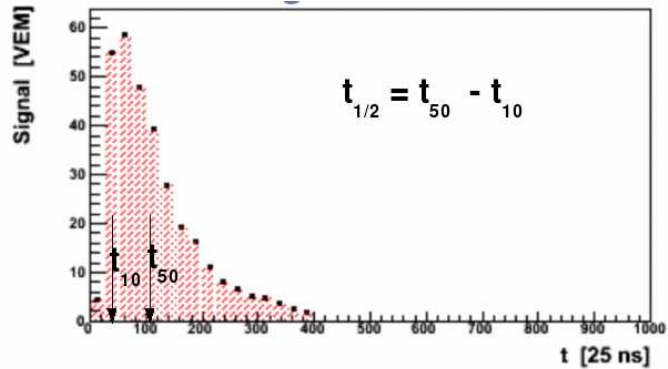,height=5cm}}
\caption {Example of a trace recorded by one PMT located in a surface detector. The arrows indicate the time interval between the arrival of 10\% and 50\% of the integrated total signal.   \label{rise_time}}
\end{center}
\end{figure}

 In addition to the $X_{max}$ measurements obtained from the FD, there are other parameters obtained with the SD that are currently being studied.  The shape of the traces observed in the PMT located in the tanks (figure \ref{rise_time}) and how they change as a function of the azimuth around the shower axis is correlated with the muon abundance, and  therefore with the primary mass composition (as discussed in section 1). A commonly used parameter to characterize the shape of the PMT traces is the rise time \cite{watson,walker1,walker2}. The rise time is the time that the traces take to collect 10\% to 50\% of the total signal (see figure \ref{rise_time}). Muons usually travel with the leading edge of the shower front, while the electromagnetic component can have some delay depending on how far from the shower axis is the sampled station. So, a smaller rise time means a larger muon abundance. It was also found that, except for vertical showers, the variation of the rise time around the shower axis is also sensitive to the mass composition (asymmetry studies). The ratio between the signal arriving earlier (within the first 600 ns) over the signal arriving later (signal arriving after 600 ns) is another measurement of the muon/electromagnetic ratio (we called this ratio the \emph{shape parameter}). For stations that are not so close to the shower axis, when a muon is detected, it generates a quite distinguishable peak. Another technique studies the peaks (or bumpiness) on the traces to estimate the muon contribution to the total signal. 

 The hybrid detection provides an enormous potential for mass composition studies that has not yet been fully exploited. Each of these parameters (rise time, asymmetry studies, shape parameter and muon abundance) may provide an independent measurement of the mass composition with independent sources of systematics, and since all these parameters were extracted from a common event, a multiparameter analysis of the mass composition may also be possible.

\section{Conclusions}

 We have presented the latest Auger results regarding the mass composition studies \cite{MU-icrc}. According to standard hadronic interaction models and the current elongation rate results, the cosmic ray composition is a mixed composition that is becoming lighter with energy up to 2 $EeV$. Above 2 $EeV$ the cosmic ray composition changes its trend and starts to become heavier with energy (see figure \ref{ER_line} plot on the right). The mass composition breaking point at 2 E$eV$ correlates with a change in the spectral index of the cosmic ray energy spectrum (figure \ref{ER_spectrum}).

 There is not a unique interpretation for the breaking point in the elongation rate data. It may be that the breaking point is not related with a mass composition change, but with a change in hadronic interaction physics. A sudden change in the hadronic interaction physics would have an effect on the expected signal at the ground for a given shower energy. This would cause a systematic error in the  reconstructed shower energy.  Therefore, an artificial break in the energy spectrum could arise due to a sudden change in the hadronic interaction physics.  

\begin{figure}[ht]
\begin{center}
\center {\epsfig{file=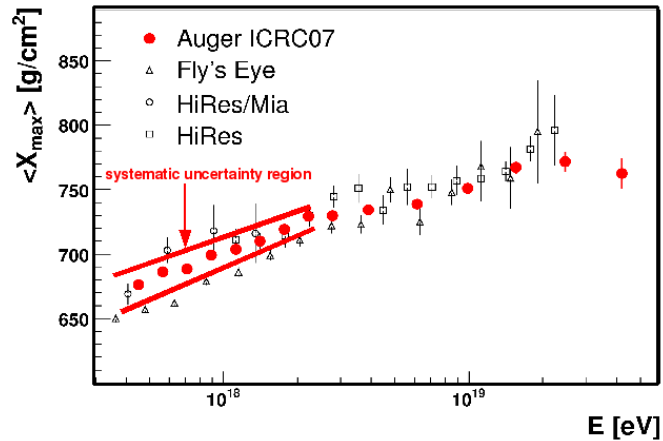,height=6cm}}
\caption {Comparing the mean $X_{max}$ measurements from the Pierre Auger Observatory with previous measurements$^{16,17,18}$.   \label{ER_others}}

\end{center}
\end{figure}

  There have been previous studies of the elongation rate by other fluorescence detectors\cite{bird,abuzayyad,abbasi}. Figure \ref{ER_others} shows that there are some systematic differences at lower energies (below 2 $EeV$) among different experiments, but for higher energies all the experiments are consistent within their uncertainties. Above 2$\times$10$^{19}$ $eV$ the other experiments run out of statistics, and Auger has still some more measurements. The last two energy bins (in the Auger data) suggest a trend towards heavier composition (according to standard hadronic interaction models). 

\section*{Acknowledgments}

 The author wish to thanks to the organizers of the XXth Rencontres de Blois 2008 ``Challenges in Particle Astrophysics'' for their invitation to present this work.


\section*{References}
\begin{thebibliography}{99}


\bibitem{sommers}P. Sommers, C. R. Physique 5, 463 (2004). 
\bibitem{risse}M. Risse, Acta Phys.Polon. {\bf B35} ,1787, (2004),  arXiv:astro-ph/0402300v1. 
\bibitem{pao}J. Abraham {\it et al} [Pierre Auger Collaboration], \Journal{\NIM} {A523}{50}{2004}.
\bibitem{bonifazi} C. Bonifazi {\it et al} [Pierre Auger Collaboration], Proc. 29$^{th}$ ICRC, , Pune, {\bf 7}, 17, (2005).

\bibitem{MU-profile}M. Unger {\it et al}, \Journal{\NIM} {A588}{433}{2008}.

 \bibitem{barbosa} H.M.J. Barbosa {\it et al}, \Journal{\APH} {22}{159}{2004}.

\bibitem{pierog} T. Pierog {\it et al} , Proc. 29$^{th}$ ICRC, , Pune, {\bf 7}, 103, (2005).  Proc. 30$^{th}$ ICRC, , Merida, (2007),  arXiv:0802.1262 [astro-ph].


\bibitem{gaisser-hillas} T. Gaisser, M. Hillas, Proc. 15$^{th}$ ICRC, , Plodiv, {\bf 8}, 353, (1977).

\bibitem{MC-simulation}S. Argiro {\it et al}, \Journal{\NIM} {A580}{1485}{2007}.
\bibitem{prado}L. Prado {\it et al}, \Journal{\NIM} {A545}{632}{2005}.
\bibitem{MU-icrc} M. Unger,  {\it et al} [Pierre Auger Collaboration],  Proc. 30$^{th}$ ICRC, , Merida, (2007), arXiv:0706.1495v1 [astro-ph].
\bibitem{roth} M. Roth,  {\it et al} [Pierre Auger Collaboration],  Proc. 30$^{th}$ ICRC, , Merida, (2007), arXiv:0706.2096v1 [astro-ph].
\bibitem{watson}  Watson, A. A., and J. G. Wilson, \Journal{\JPH} {7}{1199}{1974}.
\bibitem{walker1} Walker, R., and A. A. Watson, \Journal{\JPH} {7}{1297}{1981}.
\bibitem{walker2} Walker, R., and A. A. Watson, \Journal{\JPH} {8}{1131}{1982}.


\bibitem{bird} D.J. Bird {\it et al} [Fly's Eye Collaboration], \Journal{\PRL} {71}{3401}{1993}.
\bibitem{abuzayyad} T. Abu-Zayyad.  {\it et al} [HiRes-MIA Collaboration], \Journal{\AJ} {557}{686}{2001}.
\bibitem{abbasi} R.U Abbasi  {\it et al} [HiRes Collaboration], \Journal{\AJ} {622}{910}{2005}.
\end{thebibliography}
\end{document}